# Atomic-Scale Visualization of a Cascade of Magnetic Orders in the Layered Antiferromagnet GdTe$_3$

Arjun Raghavan[1†], Marisa Romanelli[1†], Julian May-Mann[2], Anuva Aishwarya[1], Leena Aggarwal[1], Anisha G. Singh[3,4], Maja D. Bachmann[3,4], Leslie M. Schoop[5], Eduardo Fradkin[2], Ian R. Fisher[3,4], Vidya Madhavan[1]*

[1]*Department of Physics and Materials Research Laboratory, University of Illinois Urbana-Champaign, Urbana, IL 61801, USA*

[2]*Department of Physics and Anthony J. Leggett Institute for Condensed Matter Theory, University of Illinois Urbana-Champaign, Urbana, IL 61801, USA*

[3]*Geballe Laboratory for Advanced Materials and Department of Applied Physics, Stanford University, Stanford, CA 94305, USA*

[4]*Stanford Institute for Materials and Energy Sciences, SLAC National Accelerator Laboratory, Menlo Park, CA 94025, USA*

[5]*Department of Chemistry, Princeton University, Princeton, NJ 08544, USA*

*Corresponding Author: vm1@illinois.edu

†Equal Contribution

# Abstract

GdTe$_3$ is a layered antiferromagnet which has attracted attention due to its exceptionally high mobility, distinctive unidirectional incommensurate charge density wave (CDW), superconductivity under pressure, and a cascade of magnetic transitions between 7 and 12 K, with as yet unknown order parameters. Here, we use spin-polarized scanning tunneling microscopy to directly image the charge and magnetic orders in GdTe$_3$. Below 7 K, we find a striped antiferromagnetic phase with twice the periodicity of the Gd lattice and perpendicular to the CDW. As we heat the sample, we discover a spin density wave with the same periodicity as the CDW between 7 and 12 K; the viability of this phase is supported by our Landau free energy model. Our work reveals the order parameters of the magnetic phases in GdTe$_3$ and shows how the interplay between charge and spin can generate a cascade of magnetic orders.

## Introduction

Layered antiferromagnets are rapidly garnering significant interest as a fertile set of materials to examine the interplay between magnetism and dimensionality, with promise in spintronics and twistronics applications[1–3]. Rare-earth tritellurides (RTe$_3$, R = La, Ce, Pr, Nd, Sm, Gd, Tb, Dy, Ho, Er, and Tm) are one such class of materials whose magnetic and electronic characteristics have been the focus of a large body of research for more than two decades[1,4–24]. This class of van der Waals materials hosts an interesting unidirectional incommensurate charge density wave state with magnitude $|\boldsymbol{q}_{cdw}| \approx \frac{2}{7}q_b$[6,8] (where $q_b = \frac{2\pi}{b}$; *b* is the lattice constant, space group is *Bmmb*) which is smoothly tunable by substitution of the different rare-earth elements[25]. In addition, all members of the series except LaTe$_3$ and TmTe$_3$ are magnetic, with each of the members aside from LaTe$_3$, TmTe$_3$, and ErTe$_3$ ordering antiferromagnetically above 2 K[25–28]. While Fermi surface nesting had been assigned as the primary cause of the CDW in earlier work,[6,29,30] more recent studies[7,31] indicate that the CDW formation is driven by strongly momentum-dependent electron-phonon coupling. The quasi-two-dimensional structure, consisting of RTe blocks separated by two Te square-lattice sheets on either side, is imprinted on the material's electronic and magnetic characteristics. The itinerant electrons that initiate the CDW transition are mainly confined to the Te planes. However, since any magnetism comes from the rare-earth atoms, the magnetic moments reside on the RTe block. The physical separation between the localized and itinerant electrons can lead to multiple magnetic ground states and offers a valuable opportunity to examine the interplay between the charge order and magnetic order in these materials[27,32–46].

Over the last three years, one member of the RTe$_3$ family, GdTe$_3$, has attracted significant attention for having the highest mobility of any van der Waals layered magnet, opening up possible applications in low-temperature spintronics devices as both an antiferromagnetic exchange layer as well as a highly conducting electrode in magnetic tunnel junctions[1]. The material exhibits superconductivity under pressure[47] and recent measurements have suggested that the CDW ordering, with T$_{CDW}$ = 378 K, may be unconventional, based on the angular dependence of its Raman-scattering mode[9,18,48]. Previous measurements[1,25,49] have shown a cascade of low temperature transitions in the specific heat and resistivity at ~12 K, ~10 K, and ~7 K which are also observed in magnetic susceptibility, suggesting that the system passes through a series of magnetic phases. However, despite the rapidly growing interest in this material, the exact nature of the magnetic phases in GdTe$_3$ remains unknown in part because the large absorption cross-section of Gd makes neutron scattering measurements difficult[11,40].

In this work, we use spin-polarized scanning tunneling microscopy (SP-STM) to investigate the low temperature magnetic states and the interplay between the CDW order and magnetic order in GdTe$_3$. We note here the spin orders in this paper (unless specified otherwise) henceforth refer to the in-plane (2D) projection of the 3D order, which is what we measure with STM. Fourier transforms of STM topographies obtained with spin-polarized tips show new peaks not observed with normal (spin-averaged) tips. This allows us to determine that the magnetic ground state is a striped antiferromagnetic (AFM) phase with $\boldsymbol{q}_{AFM} = \left(\pm\frac{\pi}{a}, 0\right)$. By tracking the temperature evolution of the AFM and CDW peaks in the Fourier transforms we discover an unusual intermediate temperature magnetic order: a spin density wave (SDW) state with the same wave vector as the CDW.

Our data are best explained by the existence of an intermediate phase where the bulk Gd moments are ordered antiferromagnetically along the *c*-axis. This *c*-axis AFM order magnetically polarizes the electrons that make up the CDW, leading to a "daughter" SDW order in the bulk with wave vector equal to the sum of the *c*-axis AFM and CDW wave vectors. Since the *c*-axis AFM wave vector is normal to the surface, the SDW and CDW orders have the same wave vectors on the surface, as we observe. We present a Ginzburg-Landau theory that shows how this coupling between the magnetic and charge orders can lead to the SDW phase. Finally, our measurements establish a new method for uncovering such "hidden" order (i.e. a spin order with the same wave vector as the CDW) by comparing the temperature dependence of STM data obtained with spin-averaged and spin-polarized tips (details on the tips we use are given in the Methods section).

## Results

**Spin-Polarized STM Measurements**

A schematic of the $GdTe_3$ lattice structure which consists of GdTe blocks separated by two Te square-lattice sheets is shown in Fig. 1a. The lattice constants *a* and *b* are nearly equivalent with a nominal value of 4.33 Å. A slight orthorhombicity in $GdTe_3$, as is the case for every member of the $RTe_3$ series, causes a ~0.1% difference between *a* and *b* with the CDW vector parallel to the longer direction; we assign this direction as *b*. The *c* direction lattice constant is 25.6 Å. The top view schematic in Fig. 1a shows the GdTe surface seen in our images. Only one species of atoms (either Te or Gd) is visible in most scans with a periodicity consistent with the Te/Gd lattice on the GdTe plane.

Rarely however, both species are visible as shown in the inset to Fig. 1c. Another STM group, in Lee, et al. (2023)[23], has shown similar topography as in this inset, but reported cleaving between the two Te-only planes. However, given that we typically image a lattice with periodicity consistent with either the Gd or the Te atoms from the GdTe plane and given that we can image the magnetic order, we determine it to be more likely that we are measuring the Gd atoms on the GdTe plane.

The unidirectional CDW is clearly visible as stripes in STM images taken with a metallic W tip as shown in Fig. 1c. The Fourier transform (FT) of this scan shows that the CDW wavevector is approximately 0.288 of the Bragg wavevector, which is close to a commensurate value of 0.286 ($2/7$). This suggests that the CDW in GdTe$_3$ is very close to commensurate, similar to what was seen in a previous STM study of the related compound TbTe$_3$[10]; a closer examination is shown in Supplementary Fig. 1. FTs of our scans show strong quasiparticle interference bands (Fig. 1d, Supplementary Fig. 2). The bands are notably most prominent perpendicular to the CDW direction due to the CDW gapping out parts of the Fermi surface (Fig. 1b) along the CDW direction. A linecut along the CDW direction of our Fourier transforms (Fig. 1e) shows the CDW at the previously reported position $\boldsymbol{q}_{CDW} \sim (0, \pm \frac{2}{7} q_b)$ along with the expected satellite peaks. dI/dV spectroscopy (Fig. S3) measured between -400 meV and +400 meV shows changes in slope at -210 meV and +210 meV which have been attributed to the edges of the charge density wave gap in previously published data[1,10].

Our next step is to determine the spin orders of the magnetic phases in GdTe$_3$. To do this we use magnetic Cr tips which enable us to image spin contrast (schematic in Fig. 2a). The spin-polarization of our tips is confirmed by imaging the antiferromagnetically-ordered Fe$_{1.03}$Te before obtaining measurements on GdTe$_3$, as shown in Fig. S4. Figs. 2b and 2c show a topography and the corresponding FT of GdTe$_3$ scanned with a Cr tip at a temperature of 4 K. Apart from the Bragg and CDW peaks seen in the W tip data, we observe two additional peaks whose wave vectors are independent of energy (Fig. S5), as marked by the white triangle in Fig. 2c. As these peaks, found in every case of more than five distinct Cr tip and GdTe$_3$ sample combinations (Fig. S5) are only present when scanning GdTe$_3$ with a magnetic Cr tip (comparison in Fig. S6), they must be associated with magnetic ordering in the samples. These new peaks occur at precisely half the Gd Bragg peak wave vector and appear along only one lattice direction (Fig. 2f) and not the other (Fig. 2g), indicating that the spins order into a commensurate striped antiferromagnetic (AFM) phase with $\boldsymbol{q}_{AFM} = (\pm\frac{\pi}{a}, 0)$. This spin order can be seen in real space by carrying out the inverse FT of the lattice and AFM peaks as shown in Fig. 2d. Given existing evidence from magnetization measurements[1] as well as density functional theory calculations[50] that the spins in GdTe$_3$ order in the *ab*-plane at low temperature, we conclude that the low temperature phase is an in-plane, striped AFM order as illustrated in Fig. 2e.

Interestingly, we find that $\boldsymbol{q}_{AFM}$ always appears transverse to $\boldsymbol{q}_{CDW}$. This is exemplified by measurements across a domain wall (Fig. 2h, dI/dV map across domain wall in Fig. S7) where the CDW order rotates by 90°. We note that the faint secondary stripe associated with the domain wall is not due to a tip effect and is rather a part of the structure of the domain wall and has been seen with other tips on other samples as shown in Supplementary Fig. S8. FTs of the topographies far from the domain wall show that the magnetic ordering flips by 90° across the boundary, concomitantly with the charge order

(Fig. 2h). This demonstrates directly that the magnetic and charge orders form transverse to another. In addition, the domain wall brings another surprise. A closer look at the vicinity of the domain wall (Supplementary Figs. 9-10) shows the presence of a bidirectional CDW accompanied by a bidirectional AFM order on both sides of the domain wall. This is clearly visible in the topography as well as the FTs in the vicinity of the domain wall. Indeed, linecuts of the FT of these topographies along the two directions (Supplementary Figs. 9-10) now show the CDW and AFM peaks in both directions. As we move away from the domain wall, the region of coexistence eventually gives way to the regular unidirectional CDW and AFM orders. This can be confirmed by taking similar linecuts of FTs in areas far from the domain wall, which now show only the unidirectional CDW and the corresponding unidirectional AFM peaks (Supplementary Figs. 11-12) in the perpendicular direction. This observation is further strengthened by an amplitude map of the two CDW orders which shows that the two CDWs at the domain wall have nearly the same amplitude (Supplementary Fig. 13). The observation of bidirectional CDW and AFM orders is intriguing and has a few implications. First, the presence of a bidirectional CDW suggests that the area near the domain wall may be locally strained. This agrees with recent measurements in $TbTe_3$ and $ErTe_3$[20,21] where relatively small applied uniaxial strain was shown to not only flip the CDW direction but to also create a region of coexistence of the two perpendicular CDW orders. Second, the AFM order appears robust against the gapping of the Fermi surface which should now occur in both directions due to the presence of the bidirectional CDW. This may indicate that the Fermi surface does not play a critical role in the establishment of the antiferromagnetic order.

We next move ahead to the temperature evolution of the AFM order. As shown in Figs. 3a-d, the antiferromagnetic Fourier peaks, marked by white triangles, weaken noticeably with increasing temperature; by 7.9 K, the peaks are entirely gone (FT in Fig. 3c; temperature-dependent topographies in Supplementary Fig. 14). Upon cooling the sample again, the peaks recover as shown from the 4.7 K scan Fourier transform in Fig. 3d; more extensive Fourier transforms are shown in Supplementary Fig. 15. This evolution can be visualized in the inverse Fourier transforms combining the Bragg and antiferromagnetic peaks shown in Figs. 3e-g. A plot of the intensity of the AFM peak in the FT as a function of temperature shown in Fig. 3h captures this transition. The AFM peak intensity drops below the background by 7.9 K which suggests that the Néel temperature for the striped AFM order is ~7.5 K. This is several degrees lower than the first magnetic transition seen in susceptibility data at ~11.5 K and immediately raises the question: what is the magnetic order that emerges at 11.5 K? We address this question next.

Shown in Fig. 4a is the intensity of the CDW peak ($I_{CDW}$) in the Fourier transform of our Cr tip topographies (shown in Supplementary Fig. 14) as a function of temperature. Intriguingly, we find that $I_{CDW}$ has a distinct temperature dependence only when measured with a Cr tip. As the sample is cooled below 12 K, $I_{CDW}$ increases, reaching a maximum at ~10 K, before rapidly decreasing back to the initial intensity between 7 and 8 K, as shown in Figs. 4a and c. This behavior is consistently reproducible as shown in Supplementary Fig. 16 which plots data from two more independently cleaved samples, each measured with a different Cr tip. In stark contrast, the intensity of the CDW peak in the FT of topographies taken with a non-magnetic W tip shows no systematic temperature dependence, as shown in Fig. 4b and Supplementary Figs. 17 and 18. The fact that the dI/dV spectra measured at 4.4 K, 9.9 K, and 13.9 K are nearly identical, as shown in Fig. 4d, rules out an electronic origin for this unusual temperature dependence.

## Discussion

An enhanced CDW Fourier peak intensity in an intermediate temperature range that is only observed with a spin-polarized tip indicates that there is a spin order with the same wave vector as the CDW. We henceforth refer to the spin order with the same wave vector as the CDW as the spin density wave (SDW). As the sample is cooled, the SDW sets in near 12 K and disappears between 7 K and 8 K. Below ~7.5 K, the striped AFM order described earlier emerges. We first consider the formation of the SDW near 12 K. There are several possible explanations for the formation of this SDW. It could, for example, be an independent magnetic order that forms near 12 K and coexists with the CDW. However, the fact that the SDW and CDW have the same wave vector would require fine tuning and renders this explanation unsatisfactory.

A more satisfactory explanation is that the SDW order occurs as a daughter order of the CDW order and an independent magnetic order that forms near 12 K. This other order could theoretically be a ferromagnetic order or an AFM order with a wave vector parallel to the $c$-axis, but only the formation of a $c$-axis AFM order near 12 K is consistent with bulk susceptibility and specific heat measurements[1,49]. As we show below, the combination of $c$-axis AFM and CDW orders would produce a daughter SDW order with wave vector equal to the sum of the $c$-axis AFM and CDW wave vectors. Since the $c$-axis AFM wave vector is normal to the $ab$-plane, the SDW and CDW have the same in-plane wave vector, without fine tuning. This can also be understood from the fact that, for a single *ab*-plane, the *c*-axis AFM is effectively a ferromagnetic order. This ferromagnetism imprints on the CDW, leading to a periodic modulation of the electronic spins with the same period as the CDW in each plane as illustrated in the schematic shown in Fig. 4e.

In the daughter SDW scenario, GdTe$_3$ near the 12 K transition is described by the following minimal Landau free energy,

$$F = m_c \left|S_{q^c}\right|^2 + \lambda_c \left|S_{q^c}\right|^4 + m_\rho \left|\rho_Q\right|^2 + \lambda_\rho \left|\rho_Q\right|^4 + m_{SDW} \left|S_{Q+q^c}\right|^2$$
$$+ g\, S^*_{Q+q^c} \cdot S_{q^c}\, \rho_Q + Higher - Order\ Corrections$$

where, $\rho_Q$ is the CDW order parameter with wave vector $Q \approx (0, \frac{2}{7} q_b, 0)$, $S_{q^c}$ is the *c*-axis AFM order parameter with wave vector $q^c$, and $S_{Q+q^c}$ is the SDW order parameter with wave vector $Q + q^c$. The simplest case is a period-2 *c*-axis AFM, where $q^c = (0,0,\frac{\pi}{c})$, but our conclusions are independent of the exact value of $q^c$. If the quadratic terms, $m_c$, $m_\rho$, and $m_{SDW}$, are negative (positive) the corresponding order parameter has (does not have) an expectation value. The $\lambda_c$, and $\lambda_\rho$ terms are the standard quartic Landau terms needed for stability, when $m_c$, $m_\rho$ are negative. As we shall discuss momentarily, it is not necessary to include a quartic term for $S_{Q+q^c}$ in our immediate analysis, and we shall therefore leave it implicit in the "Higher-order Corrections." We additionally have the non-trivial tri-linear $g$ term, which couples the CDW, c-axis AFM, and SDW order parameters. Physically, this term describes how the wavevector $Q$ charge density wave can imprint on the wavevector $q^c$ spin density wave, leading to additional spatial oscillations of the spins with wavevector $Q + q^c$. We are interested in the situation where the c-axis AFM order develops below $T_c \sim 12\,K$, which corresponds to setting $m_c = \alpha\,(T - T_c)$ for some $\alpha > 0$. Since the CDW order develops at much higher temperatures, we use a constant value of $m_\rho, < 0$ such that $\langle \rho_Q \rangle \neq 0$. We also set $m_{SDW} > 0$, such that an independent $S_{Q+q^c}$ SDW order does not form near $12\,K$. Because we are only considering $m_{SDW} > 0$ the Landau free energy is stable without including a quartic term for $S_{Q+q^c}$. We have not included all symmetry-allowed terms here; notably, we are missing biquadratic terms that promote the coexistence of different ordering. However, this minimal free energy is sufficient to capture how a daughter SDW order can arise from the combination of *c*-axis AFM and CDW orders.

For $T < T_c$, there is c-axis AFM order, $\langle S_{q^c} \rangle \neq 0$. Because of the tri-linear $g$ term, the free energy is minimized when the SDW order also has an expectation value, $\langle S_{Q+q^c} \rangle = -\frac{g}{m_{SDW}} \langle S_{q^c} \rangle \langle \rho_Q \rangle$ for $T < T_c$. The combination of CDW order, $\langle \rho_Q \rangle \neq 0$, and *c*-axis AFM order, $\langle S_{q^c} \rangle \neq 0$, therefore produces a daughter SDW order, $\langle S_{Q+q^c} \rangle \neq 0$, as discussed above. It is worth stating explicitly that the above Landau free energy only describes the system near the 12 K transition, and does not capture the magnetic transition that occurs near 7.5 K.

We next turn our attention to the disappearance of the SDW and formation of the striped AFM that occurs near 7.5 K. A simple scenario to explain the low temperature phase is that the *c*-axis AFM present between 7.5 K and 12 K rotates to the *ab*-plane below 7 K where it becomes the striped AFM that we observe. The interplay between the low temperature and the intermediate phase orderings is presumably due to competing interactions. Alternatively, it could be the case that the *c*-axis and striped AFM phases are separated by a small paramagnetic phase. This would mean that there is a reentrant phase transition near 7.5 K that is likely driven by entropy effects[44,45]. Finally, the *c*-axis and striped AFM phases could simply be directly connected via a first-order phase transition[1].

Using spin-polarized STM, we have thus identified the in-plane low temperature magnetic phases of GdTe$_3$. Our data show that there are two distinct magnetic orders – an emergent spin density wave with wave vector $\bm{q}_{SDW} = \bm{q}_{CDW} \sim (0, \pm\frac{2}{7}q_b)$ that sets in at ~12 K and likely coexists with a *c-axis* AFM order, and a striped antiferromagnetic order with wave vector $\bm{q}_{AFM} = (\pm\frac{\pi}{a}, 0)$ that sets in at ~7.5 K. We propose that the SDW occurs as a daughter order of an intermediate temperature *c*-axis AFM phase and the CDW, while the striped AFM is the actual low temperature ground state. Through this work, we have also demonstrated how comparing the temperature dependence of Fourier transform intensities measured with spin-polarized and spin-averaged STM tips enables the detection of hidden spin orders that may have the same periodicity as different orders in the system. Future experiments with SP-STM on the rare-earth tritellurides in the presence of an in-plane magnetic field could provide further evidence elucidating the complex interplay between magnetic and charge orders in this distinctive class of materials.

## Methods

**Crystal Growth, STM Measurements, and Data Processing**

GdTe$_3$ single crystals were grown by a self-flux technique, as reported previously[25,29,51]; crystals used in this work have approximate dimensions of 2 mm × 2 mm. STM measurements were performed after cleaving single crystals of $Fe_{1.03}Te$ and $GdTe_3$ at 77 K. W tip wires of 250 μm diameter and Cr tip wires of 300 μm diameter are used for measurements. W tips are electrochemically etched in an NaOH solution, while Cr tips are electrochemically etched using $H_2SO_4$. Additional data with a W tip is shown in Supplementary Figs. 17, 18, and 19. Both types of tips are annealed in vacuum prior to measurements. Cr tips are used for the spin-polarized measurements described, and are calibrated on antiferromagnetic $Fe_{1.03}Te$ single crystals, as shown in a number of previous works[26,38,52]. Example calibration is shown in Fig. S4 and examples of the robustness of maintaining spin-polarization in Cr tips while changing samples are shown in Supplementary Fig. 20. For scanning tunneling spectroscopy, a Stanford Research Systems SR830 lock-in amplifier is used, with dI/dV spectra obtained at a lock-in frequency of 913 Hz. To perform temperature dependence measurements, we employ Joule heating with a Cu coil; the temperatures are set and maintained with a Lakeshore 340 controller. Fourier peak intensities are extracted by taking the average intensity value of a 3 × 3-pixel area centered on the peak. Background intensity is determined by taking the average intensity of a 3 × 3-pixel area in a region of the Fourier transform far from any peaks. The intensity of the centermost pixels of Fourier transforms is suppressed so they don't dominate the color contrast. In the limited cases where Fourier transforms have been mirror-symmetrized, we have explicitly noted this step in the caption; while features of the Fourier transform are occasionally clearer visually after this step, the Bragg peaks, CDW peaks, and quasiparticle interference patterns are visible even without this processing and the mirror-symmetrization does not affect our conclusions.

## Acknowledgments

We acknowledge the National Science Foundation for supporting STM measurements through Grant No. DMR-2003784. In addition, V.M. acknowledges partial support from the Gordon and Betty Moore Foundation's EPiQS Initiative through Grant No. GBMF9465. M.R. and L.M.S. were supported by the Air Force Office of Scientific Research (AFOSR) under Grant No. FA9550-23-1-0635. I.R.F., A.G.S., and M.D.B. (crystal growth and characterization) were supported by the Department of Energy, Office of Basic Energy Sciences, under contract DE-AC02-76SF00515. J.M-M. is supported by the National Science Foundation Graduate Research Fellowship Program under Grant No. DGE-1746047. This work was also supported in part by the National Science Foundation Grant No. DMR-2225920 at the University of Illinois (E.F.).

## Data Availability

Raw data presented in this work are available through the Illinois Data Bank at https://databank.illinois.edu/datasets/IDB-4638513.

## Author Contributions

A.R., M.R., A.A., and L.A. performed STM measurements and analyzed data. Single crystals were provided by A.G.S., M.D.B., and I.R.F. J.M-M., E.F., and L.M.S. provided theoretical input and helped with interpretation of the data. A.R., M.R., J.M-M., and V.M. wrote the paper, with contributions from all authors.

## Competing Interests

The authors declare no competing interests.

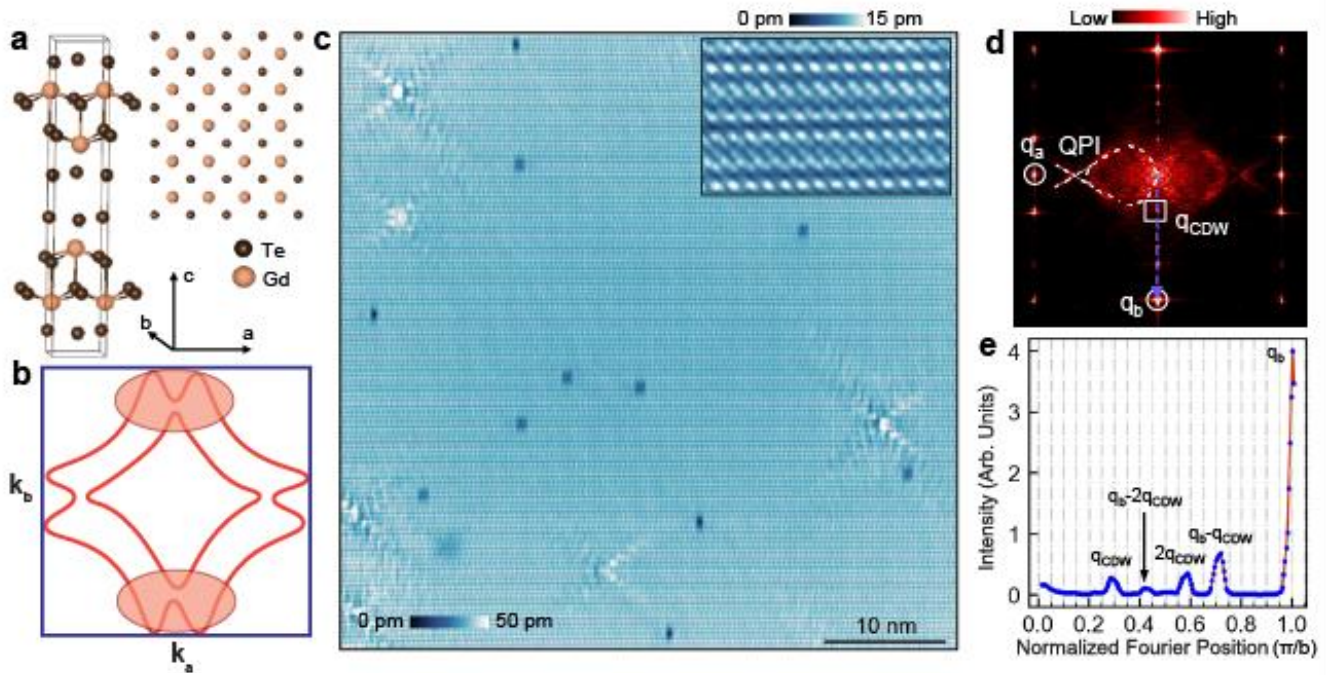

**Fig. 1. Crystal Structure, Fermi Surface, Topography, and dI/dV Spectra. a** Crystal structure of GdTe$_3$, with a top view of the GdTe surface shown to the right. **b** Sketch of Fermi surface. The shaded areas are a schematic representation of Fermi surface gapping associated with the unidirectional CDW. **c** Large-area (800 Å x 800 Å, $V_{Bias}$ = -120 mV, $I_{Set}$ = 240 pA) and atomic-resolution (top right inset, 6 Å × 3.75 Å, $V_{Bias}$ = -400 mV, $I_{Set}$ = 120 pA) topographies of cleaved GdTe$_3$. All topographies in this manuscript are shown with a linear plane subtracted. **d** FT of the large-area W tip scan in **c**, with the CDW and Bragg peaks labeled as well as quasiparticle interference (QPI) bands observed. The QPI bands in the CDW direction (along $q_b$) are gapped out. This FT has been mirror-symmetrized and had a first nearest-neighbor averaging applied. **e** Line profile along the blue arrow shown in **d**. All data shown in this figure were taken at 4 K.

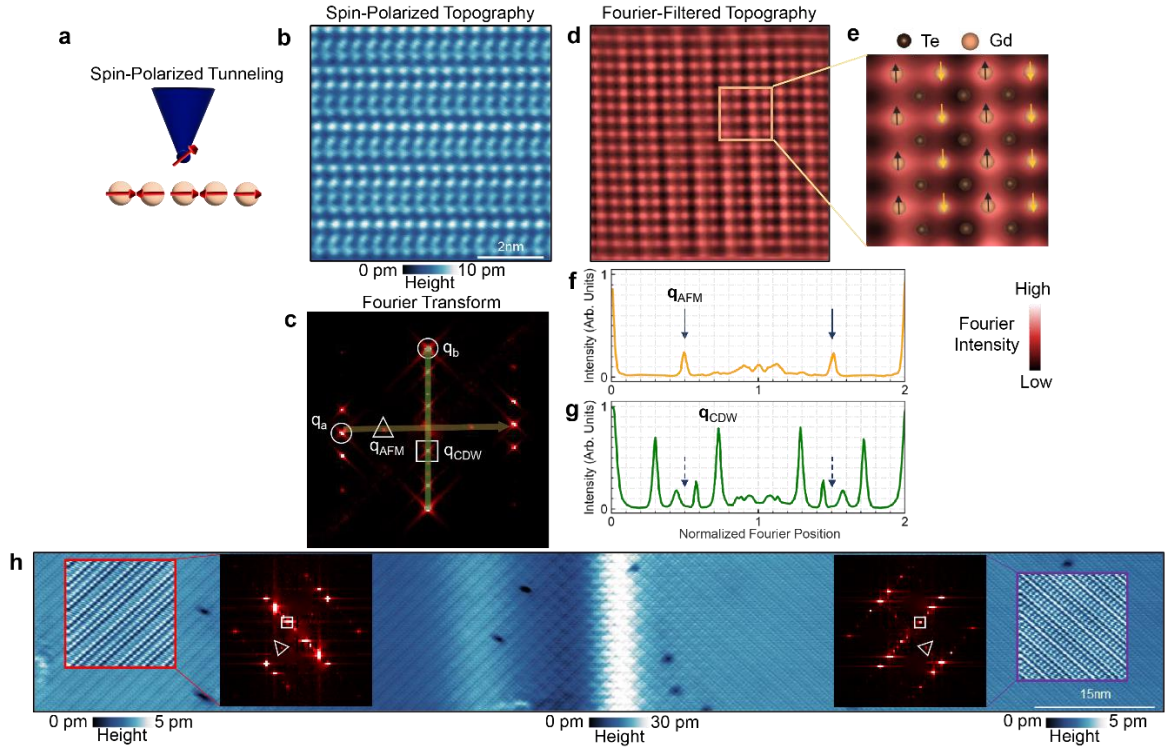

**Fig. 2. Striped Antiferromagnetic Ordering in GdTe$_3$. a** Schematic of spin-polarized STM. **b** 80 Å × 80 Å topography of GdTe$_3$ with a spin-polarized Cr tip (V$_{Bias}$ = -200 mV, I$_{Set}$ = 280 pA). **c** FT of the GdTe$_3$ scan with a spin-polarized Cr tip in **b**, showing Bragg peaks, CDW peaks, and an extra pair of peaks transverse to the CDW, arising due to commensurate striped antiferromagnetic (AFM) ordering. A first nearest-neighbor average has been applied to the FT. **d** Fourier-filtered scan by combining Bragg Fourier peaks, showing a square lattice, with striped AFM order; the Fourier transform used for filtering is that shown in **c**. **e** A zoom-in of the boxed region in **d** with positions of Te and Gd atoms displayed, along with a schematic of the spin order. **f** Line profile along yellow line in **c**, showing the AFM peaks at exactly half the Bragg peak distance; y-axis normalization is to the maximum Fourier amplitude. **g** Line profile along green line in **c**, displaying CDW and satellite peaks, with y-axis normalization to the maximum Fourier amplitude. Notably, there is no peak at half the Bragg peak distance as found in **f**; hence, the antiferromagnetism is a commensurate stripe order. **h** 1500 Å × 300 Å topography (V$_{Bias}$ = -60 mV, I$_{Set}$ = 60 pA) across a CDW domain wall in GdTe$_3$. Topography insets show high resolution 100 Å × 100 Å scans on either side of the domain wall; parameters are V$_{Bias}$ = -60 mV, I$_{Set}$ = 60 pA. FTs of the inset topographies show the CDW direction rotating, along with a simultaneous rotation of the transverse antiferromagnetic peaks. This shows that the CDW and commensurate AFM ordering in GdTe$_3$ always form transverse to one another. All data shown in this figure were taken at 4 K.

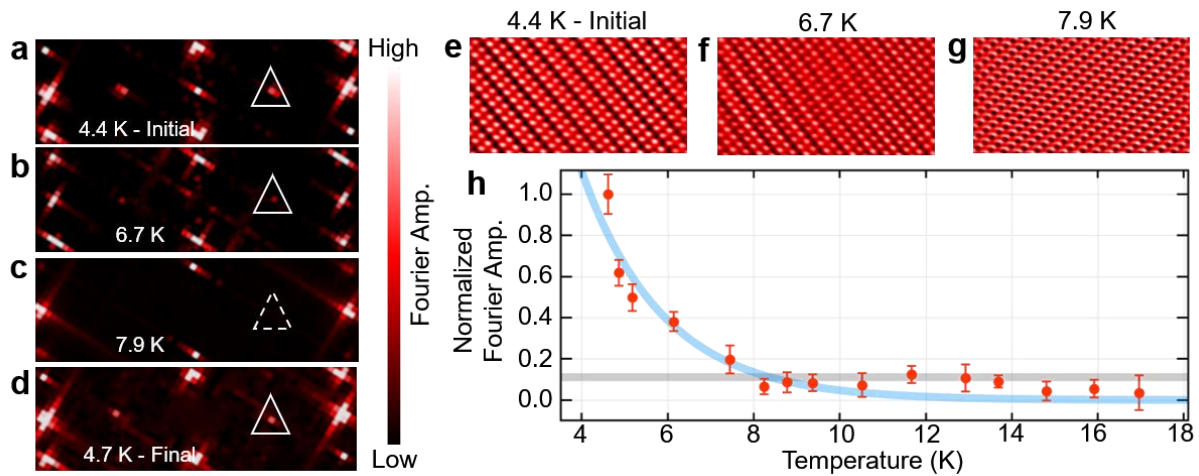

**Fig. 3. Temperature Dependence of Antiferromagnetism on GdTe$_3$. a-d** FTs of AFM Fourier peaks from Cr tip scans at $V_{Bias}$ = -400 mV, $I_{Set}$ = 60 pA on GdTe$_3$ (topographies shown in Supplementary Fig. 14) show strong peaks at 4.4 K, a weaker peak at 6.7 K, and the absence of AFM peaks when heated to 7.9 K and above; cooling back down shows clear AFM peaks at 4.7 K. The displayed FTs have been first nearest-neighbor averaged. **e-g** Inverse FTs of Bragg peaks and AFM peaks from **a-c** show the decay of the magnetic order upon heating; the AFM inverse Fourier intensity is multiplied by 5 to show the contrast more clearly. **h** Plot of the temperature dependence of AFM Fourier peak intensity, normalized to the maximum intensity value. Blue curve is a guide to the eye and light gray line is the average background intensity level, with its thickness corresponding to the standard deviation of the background Fourier intensity.

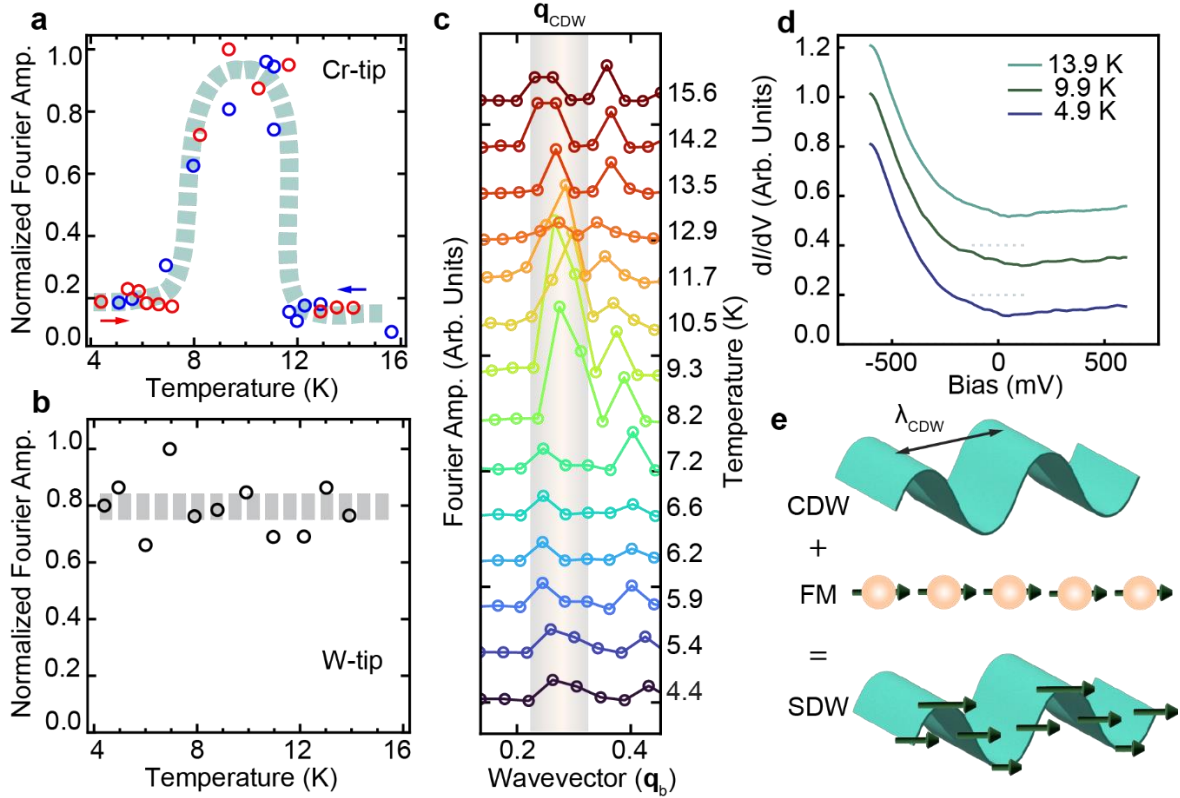

**Fig. 4. Evolution of Magnetic Ordering with Temperature. a** CDW peak intensities from FTs of scans on GdTe$_3$ taken with a spin-polarized Cr tip at $V_{Bias}$ = -400 mV, $I_{Set}$ = 60 pA, as a function of temperature (topographies shown in Supplementary Fig. 14); y-axis normalization is to the maximum Fourier amplitude in the Cr tip run. Three distinct regimes are observed: a low-temperature region below 7 K in which the intensities are constant, an intermediate temperature phase between 7 K and 12 K where the intensities are considerably larger, and a high temperature region in which the intensities return to approximately the same, constant intensity as found below 7 K. Red points were taken on the temperature upsweep, while blue points were taken on the down sweep. **b** CDW peak intensities from FTs of GdTe$_3$ scans taken with a spin-degenerate W tip at $V_{Bias}$ = -240 mV, $I_{Set}$ = 240 pA, as a function of temperature (topographies shown in Supplementary Fig. 18), with y-axis normalization to the maximum Fourier amplitude in the W tip run. The intensities stay approximately constant, and any changes are random. **c** Linecuts along the CDW direction from FTs of the temperature-dependence, cropped to show the area around the CDW peak (highlighted by the shaded region). The CDW peak intensity significantly increases in the intermediate temperature regime. **d** Representative dI/dV spectra taken with a W tip on GdTe$_3$ in each of the 3 temperature regimes; the electronic density of states is nearly identical, providing further evidence that the pattern in **a** is due to magnetic ordering. Setpoint for dI/dV spectra are $V_{Bias}$ = -600 mV and $I_{Set}$ = 60 pA. **e** Schematic illustrating how the interaction between CDW order and ferromagnetic (FM) order would result in a daughter spin density wave order, represented by the spins of varying magnitudes in the bottom row.

# Supplementary Information

# Atomic-Scale Visualization of a Cascade of Magnetic Orders in the Layered Antiferromagnet GdTe$_3$

Arjun Raghavan[1†], Marisa Romanelli[1†], Julian May-Mann[2], Anuva Aishwarya[1], Leena Aggarwal[1], Anisha G. Singh[3,4], Maja D. Bachmann[3,4], Leslie M. Schoop[5], Eduardo Fradkin[2], Ian R. Fisher[3,4], Vidya Madhavan[1]*

[1]*Department of Physics and Materials Research Laboratory, University of Illinois Urbana-Champaign, Urbana, IL 61801, USA*

[2]*Department of Physics and Anthony J. Leggett Institute for Condensed Matter Theory, University of Illinois Urbana-Champaign, Urbana, IL 61801, USA*

[3]*Geballe Laboratory for Advanced Materials and Department of Applied Physics, Stanford University, Stanford, CA 94305, USA*

[4]*Stanford Institute for Materials and Energy Sciences, SLAC National Accelerator Laboratory, Menlo Park, CA 94025, USA*

[5]*Department of Chemistry, Princeton University, Princeton, NJ 08544, USA*

*Corresponding Author: vm1@illinois.edu

†Equal Contribution

**Supplementary Fig. 1 | Charge Density Wave Periodicity in GdTe$_3$**

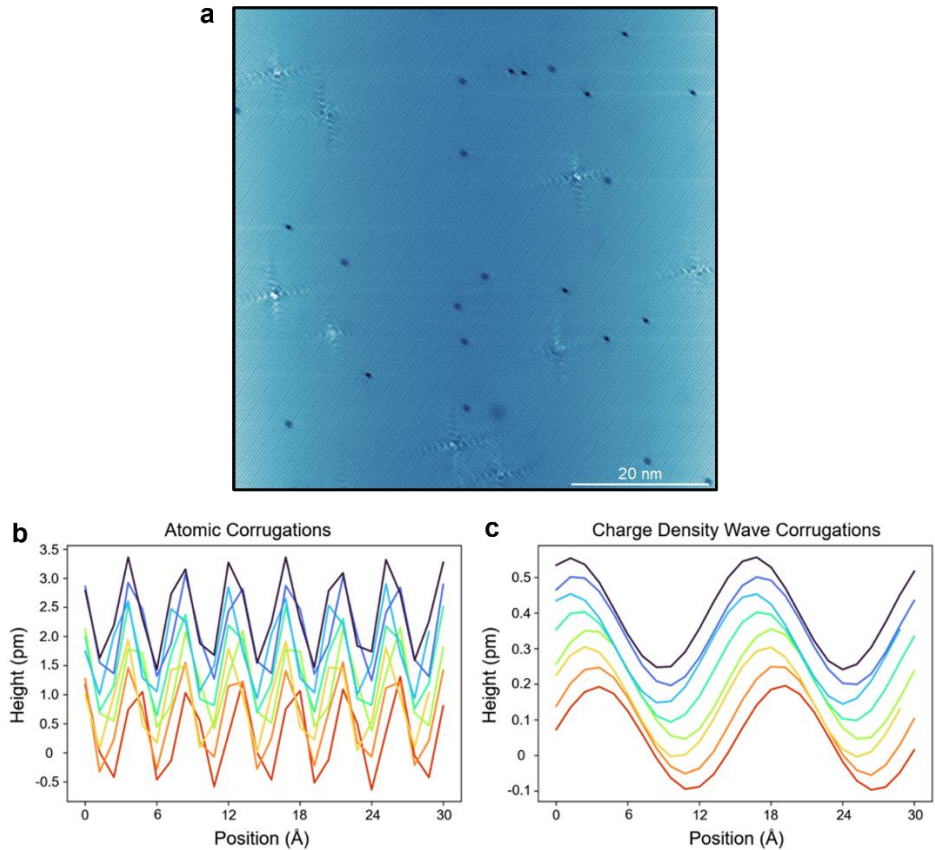

**a** 700 Å × 700 Å scan of GdTe$_3$ with a W tip. Scan parameters are $V_{Bias}$ = -240 mV and $I_{Set}$ = 240 pA. **b** Linecut across a Fourier-filtered topography shown in **a**; the lines are broken after every 7 atoms and plotted on top of each other with small offsets. Fourier filtering is of the Gd Bragg peaks. **c** Linecut across the same scan in GdTe$_3$ as in **a**, but with Fourier filtering of the charge density wave peaks. The lines are broken at the same spatial intervals as those in **a**, and show only a very small shift in each subsequent line, indicating that the CDW periodicity appears to be very close to $(^7/_2)b$, within approximately 0.5%. All data shown in this figure was taken at 4 K, $V_{Bias}$ = -240 mV, and $I_{Set}$ = 240 pA.

**Supplementary Fig. 2 | Bias-Dependent Fourier Transforms from dI/dV Map**

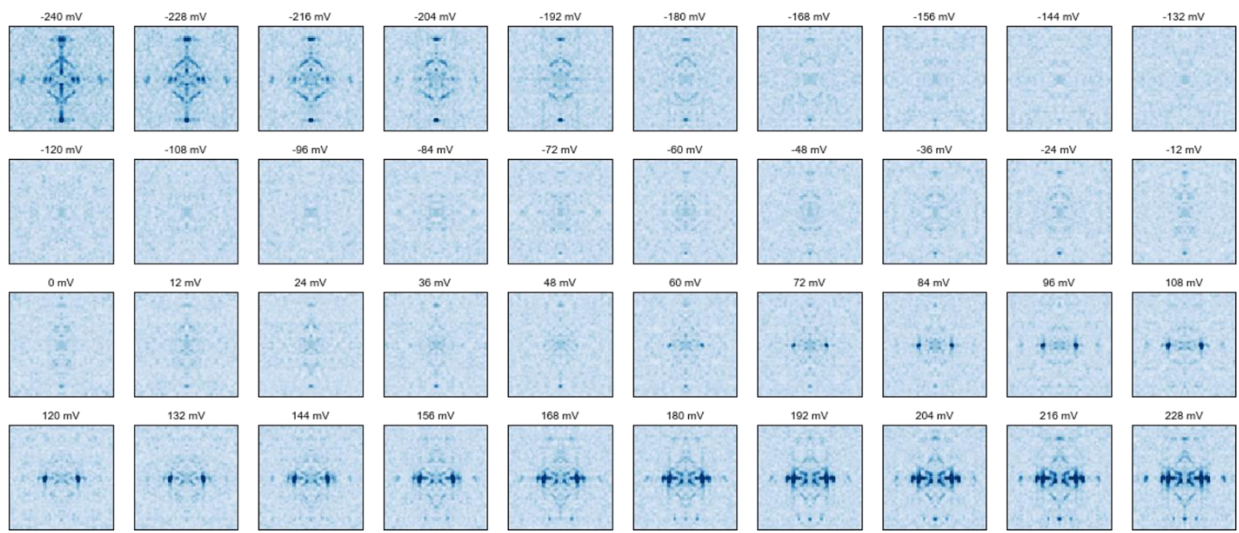

Fourier transforms taken at each bias from a dI/dV map taken with a metallic W tip show dispersing QPI bands. All Fourier transforms in this figure have been mirror symmetrized and had first nearest-neighbor averaging applied. The map was taken at 4 K with a setpoint of 60 pA at -240 mV, with a modulation of 4 mV.

**Supplementary Fig. 3 | dI/dV Spectroscopy Linecut on GdTe$_3$**

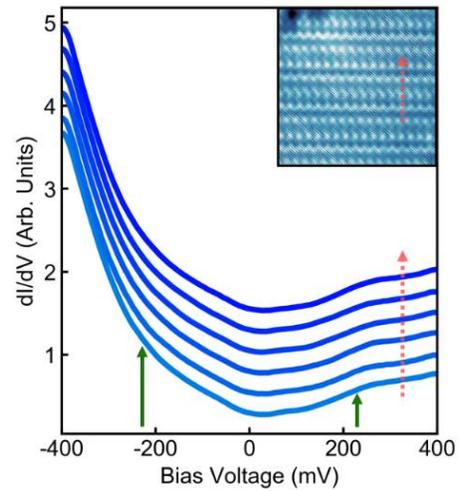

dI/dV spectra ($V_{Bias}$ = -400 mV, $I_{Set}$ = 260 pA) along arrow in 70 Å × 70 Å scan shown in inset, displaying strong electronic homogeneity across CDW stripes; green arrows indicate changes in slope of spectra at -210 mV and +210 mV, corresponding to the expected energy values of the CDW gap edges. All data shown in this figure was taken at 4 K.

**Supplementary Fig. 4 | Calibration of Spin-Polarized Cr Tips on $Fe_{1.03}Te$**

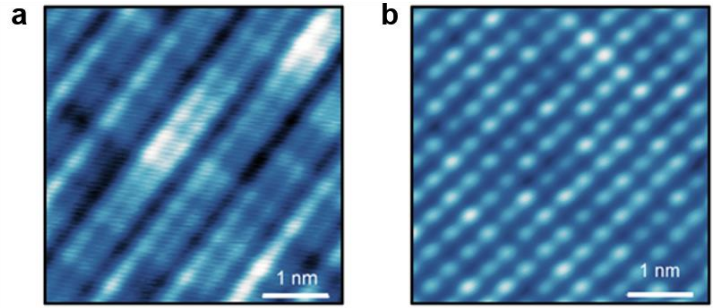

**a** Topography scan of $Fe_{1.03}Te$ with a Cr tip, showing a square lattice with additional stripe modulation corresponding to bicollinear antiferromagnetism; spin-polarized tips are prepared by training on $Fe_{1.03}Te$ until a clear magnetic ordering is visible, as in the image shown ($V_{Bias}$ = 380 mV, $I_{Set}$ = 50 pA). **b** Comparison scan ($V_{Bias}$ = -120 mV, $I_{Set}$ = 60 pA) of $Fe_{1.03}Te$ measured with a W tip, showing a square lattice without the additional stripe modulation found in **a**. Scan sizes of the two topographies displayed in this figure are 42.5 Å × 42.5 Å. All data shown in this figure was taken at 4 K.

**Supplementary Fig. 5 | Repeatability of Detecting Antiferromagnetic Ordering in GdTe$_3$**

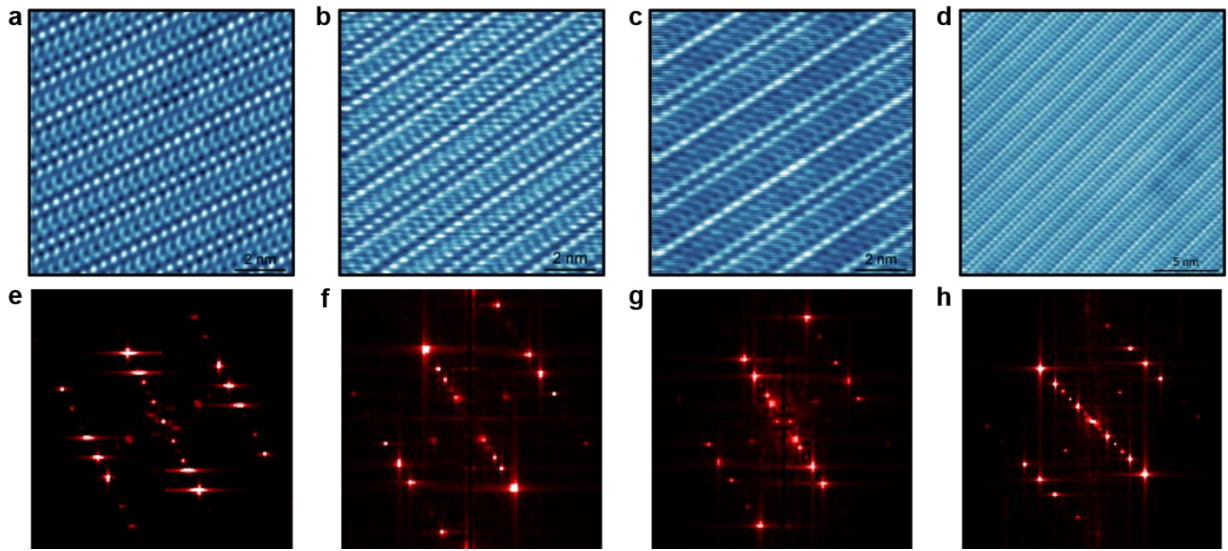

**a-h** Topography scans and Fourier transforms from 4 different GdTe$_3$ samples measured by 4 different spin-polarized Cr-tips showing a pair of peaks transverse to the CDW indicating collinear antiferromagnetic ordering. Scan parameters are: 100 Å × 100 Å with $V_{Bias}$ = 60 mV and $I_{Set}$ = 480 pA in **a** (Fourier transform is **e**), 100 Å × 100 Å with $V_{Bias}$ = -400 mV and $I_{Set}$ = 60 pA in **b** (Fourier transform is **f**), 100 Å × 100 Å with $V_{Bias}$ = -60 mV and $I_{Set}$ = 60 pA in **c** (Fourier transform is **g**), and 200 Å × 200 Å with $V_{Bias}$ = -180 mV and $I_{Set}$ = 120 pA in **d** (Fourier transform is **h**). It is evident that the antiferromagnetism can be detected by spin-polarized STM scans both inside and outside the CDW gap, and both above and below the Fermi energy. All data shown in this figure was taken at 4 K.

**Supplementary Fig. 6 | W-Tip and Cr-Tip Scans Fourier Transform Comparison**

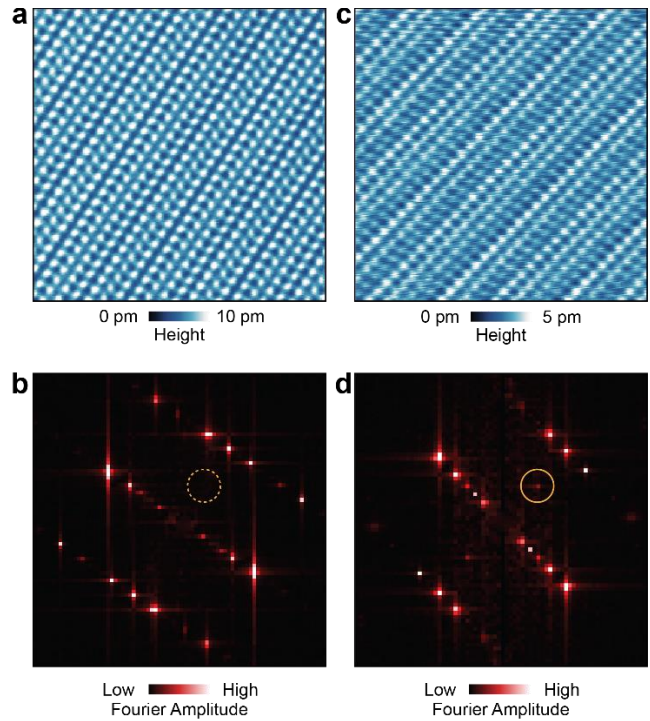

**a** 100 Å × 100 Å scan of GdTe$_3$ with a W tip at $V_{Bias}$ = -120 mV and $I_{Set}$ = 240 pA. **b** Fourier transform of scan in **a** shows no extra peaks (marked with a dashed yellow circle) aside from Bragg peaks, CDW peaks, and satellite peaks. **c** 100 Å × 100 Å scan of GdTe$_3$ with a Cr tip at $V_{Bias}$ = -100 mV and $I_{Set}$ = 60 pA. **d** Fourier transform of scan in **c** shows two clear extra peaks (position marked by yellow circle) halfway between the center and the Gd Bragg peaks perpendicular to the CDW direction. All data shown in this figure was taken at 4 K.

## Supplementary Fig. 7 | dI/dV Spectroscopy of a Domain Wall in GdTe$_3$

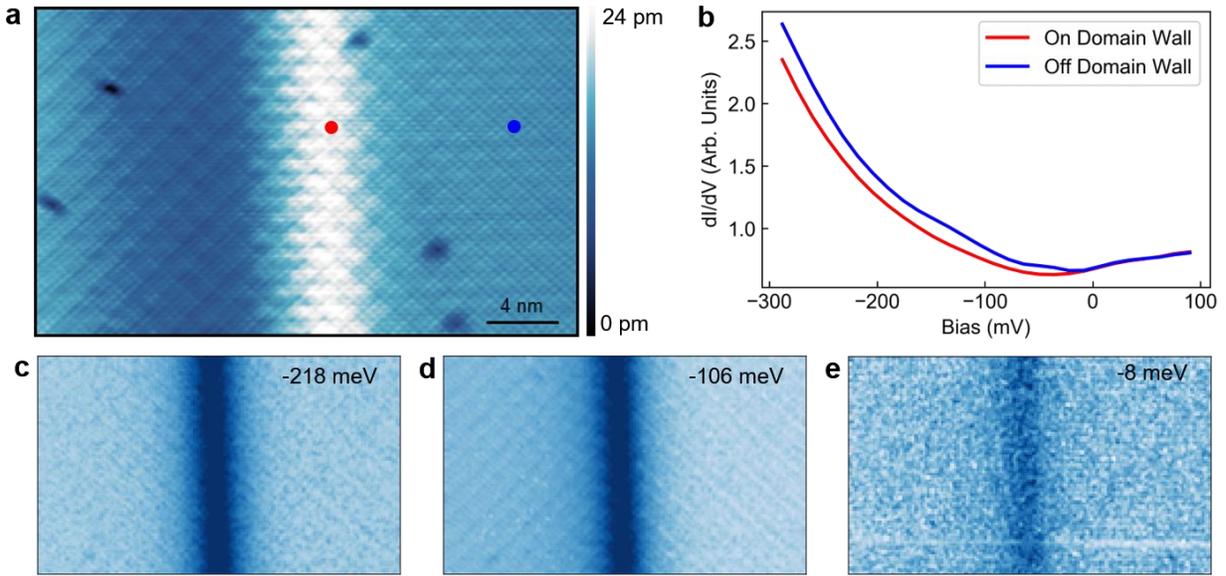

**a** 300 Å × 180 Å scan ($V_{Bias}$ = -60 mV, $I_{Set}$ = 60 pA) of a domain wall, showing that a weakening second CDW is present on both sides of the domain wall. **b** dI/dV spectra ($V_{Bias}$ = 90 mV, $I_{Set}$ = 60 pA) on and away from the domain wall. **c-e** Slices (300 Å × 180 Å) of a dI/dV map with a setpoint of $V_{Bias}$ = 90 mV and $I_{Set}$ = 60 pA taken across the domain wall. All data shown in this figure was taken at 4 K.

**Supplementary Fig. 8 | Comparison of Domain Walls on GdTe$_3$**

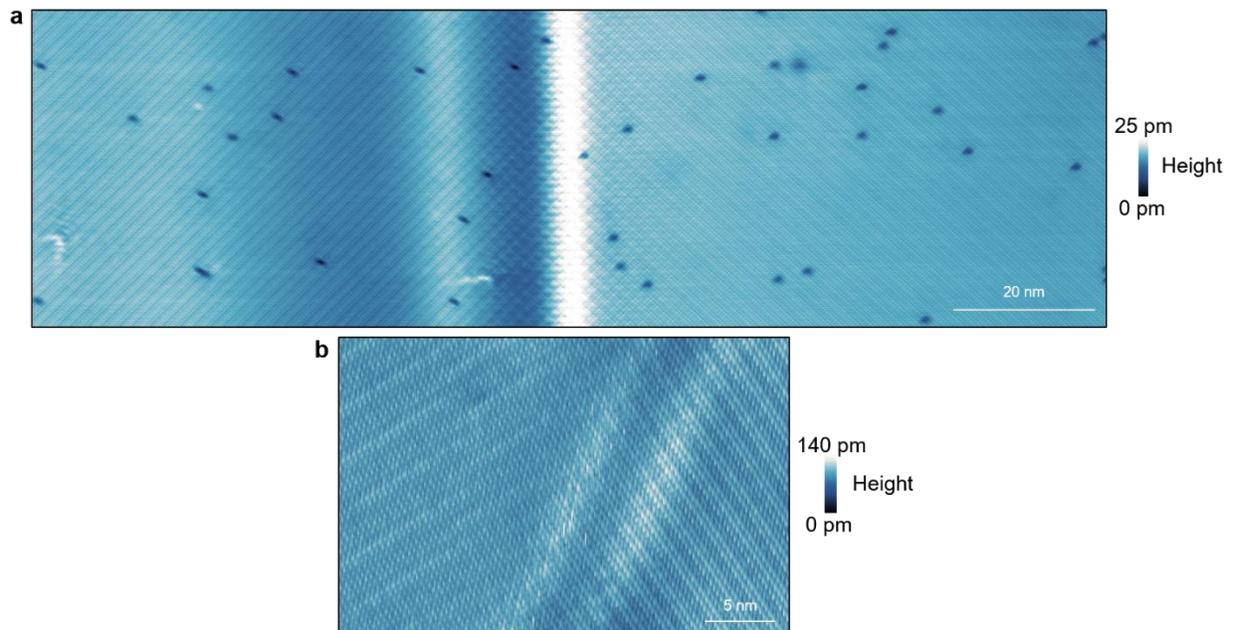

**a** Cr tip scan of a domain wall on GdTe$_3$. Scan size is 1500 Å × 450 Å; $V_{Bias}$ = -60 mV, $I_{Set}$ = 60 pA. **b** W tip scan of a different domain wall. Scan size is 300 Å × 200 Å; $V_{Bias}$ = -300 mV, $I_{Set}$ = 60 pA. Both domain walls have more than a single bright feature. The features are most likely intrinsic to the sample and not due to a double-tip effect since they appear on different samples measured with different tips and since defects in **a** appear as single dark spots. All data shown in this figure was taken at 4 K.

**Supplementary Fig. 9 | Antiferromagnetic Peaks Near Domain Wall (Left Side)**

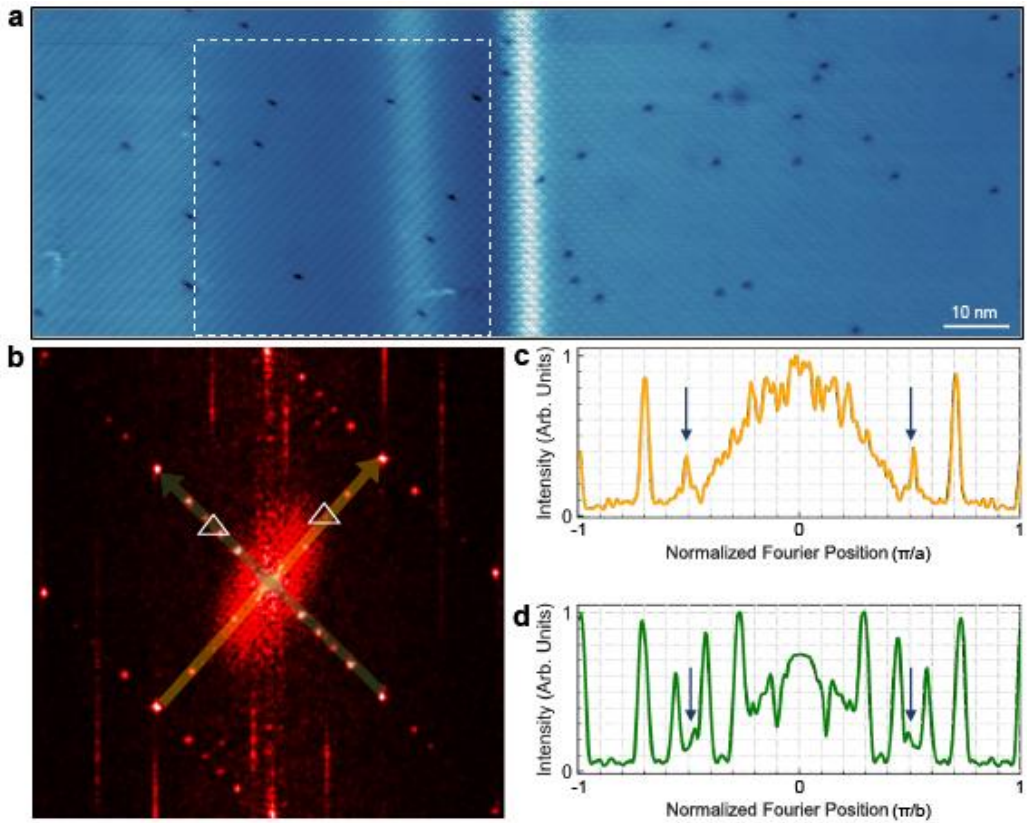

**a** Topography of GdTe$_3$ domain wall region with Cr tip ($V_{Bias}$ = -60 mV, $I_{Set}$ = 60 pA); scan size is 1500 Å × 500 Å. **b** Fourier transform of area in topography marked by white square in **a**; antiferromagnetic (AFM) peaks along both directions are visible and are marked by triangles. **c** Linecut along orange arrow in **b**; AFM and CDW peaks are marked and normalization is to the maximum Fourier amplitude. **d** Linecut along green arrow in **b**; AFM and CDW peaks are marked and normalization is to the maximum Fourier amplitude. All data shown in this figure was taken at 4 K.

**Supplementary Fig. 10 | Antiferromagnetic Peaks Near Domain Wall (Right Side)**

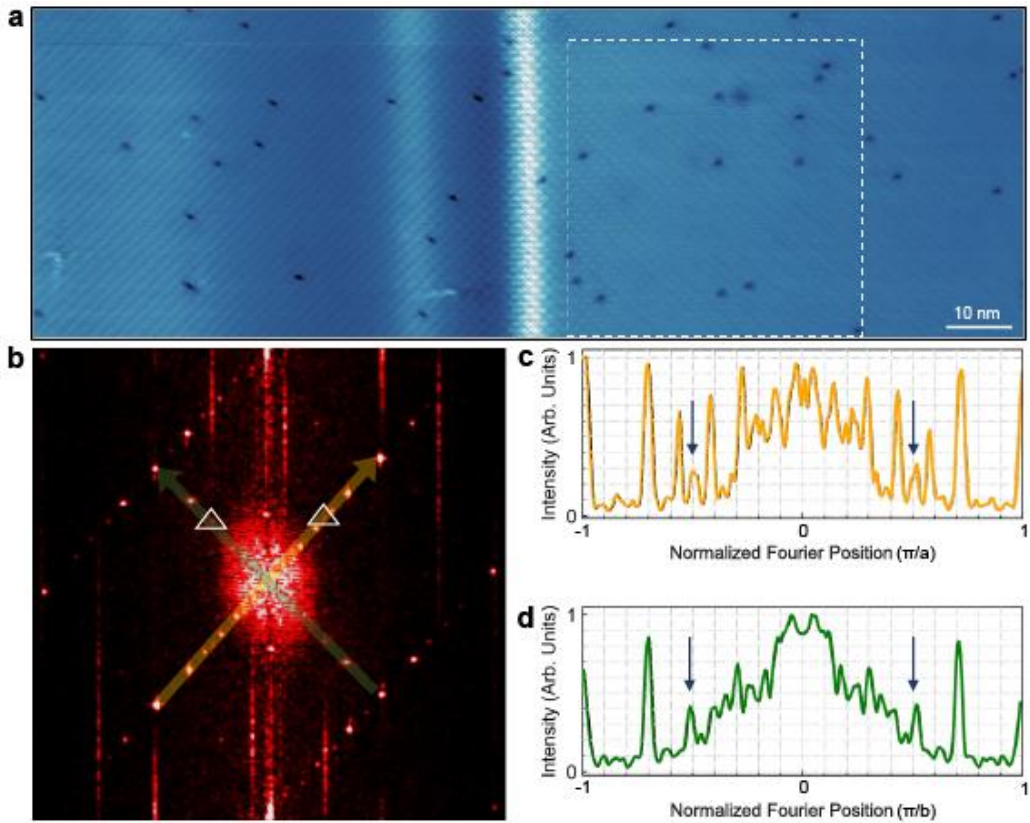

**a** Same topography of GdTe$_3$ domain wall region with Cr tip as in Fig. S9 ($V_{Bias}$ = -60 mV, $I_{Set}$ = 60 pA); scan size is 1500 Å × 500 Å. **b** Fourier transform of area in topography marked by white square in **a**; antiferromagnetic (AFM) peaks along both directions are visible and are marked by triangles. **c** Linecut along orange arrow in **b**; AFM and CDW peaks are marked and y-axis normalization is to the maximum Fourier amplitude. **d** Linecut along green arrow in **b**; AFM and CDW peaks are marked and y-axis normalization is to the maximum Fourier amplitude. All data shown in this figure was taken at 4 K.

**Supplementary Fig. 11 | Antiferromagnetic Peaks Away from Left of Domain Wall**

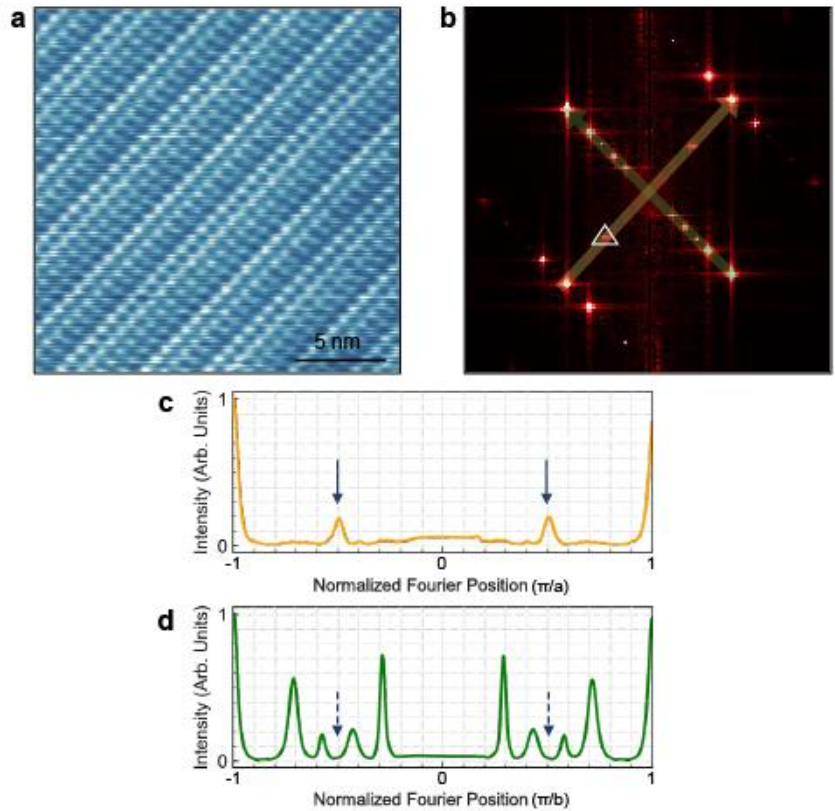

**a** Cr tip topography far from left of domain wall (200 Å x 200 Å, $V_{Bias}$ = -240 mV, $I_{Set}$ = 180 pA). **b** Fourier transform of area in topography in **a**; antiferromagnetic (AFM) peaks along only one direction is visible and is marked by a triangle. **c** Linecut along orange arrow in **b**; AFM and CDW peaks are marked and normalization y-axis normalization is to the maximum Fourier amplitude. **d** Linecut along green arrow in **b**; dashed blue arrows indicate absence of AFM peaks. AFM and CDW positions are marked and y-axis normalization is to the maximum Fourier amplitude. All data shown in this figure was taken at 4 K.

**Supplementary Fig. 12 | Antiferromagnetic Peaks Away from Right of Domain Wall**

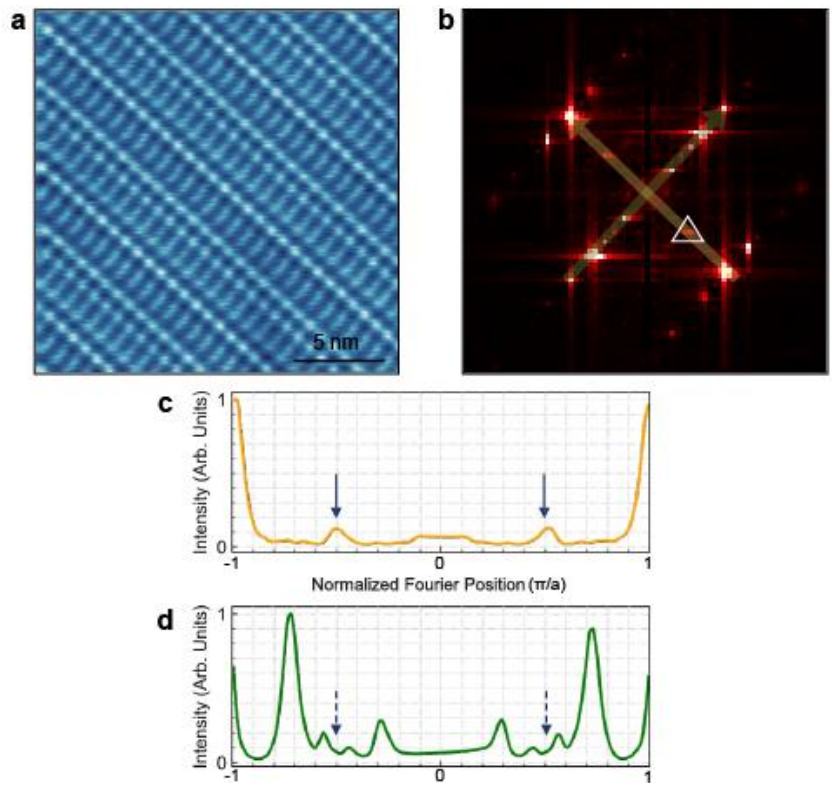

**a** Cr tip topography far from right of domain wall (200 Å x 200 Å, $V_{Bias}$ = -60 mV, $I_{Set}$ = 60 pA). **b** Fourier transform of area in topography in **a**; antiferromagnetic (AFM) peaks along only one direction is visible and is marked by a triangle. **c** Linecut along orange arrow in **b**; solid blue arrows indicate AFM peaks and y-axis normalization is to the maximum Fourier amplitude. **d** Linecut along green arrow in **b**; dashed blue arrows indicate absence of AFM peaks. AFM and CDW positions are marked and y-axis normalization is to the maximum Fourier amplitude. All data shown in this figure was taken at 4 K.

**Supplementary Fig. 13 | CDW Amplitudes Near Domain Wall**

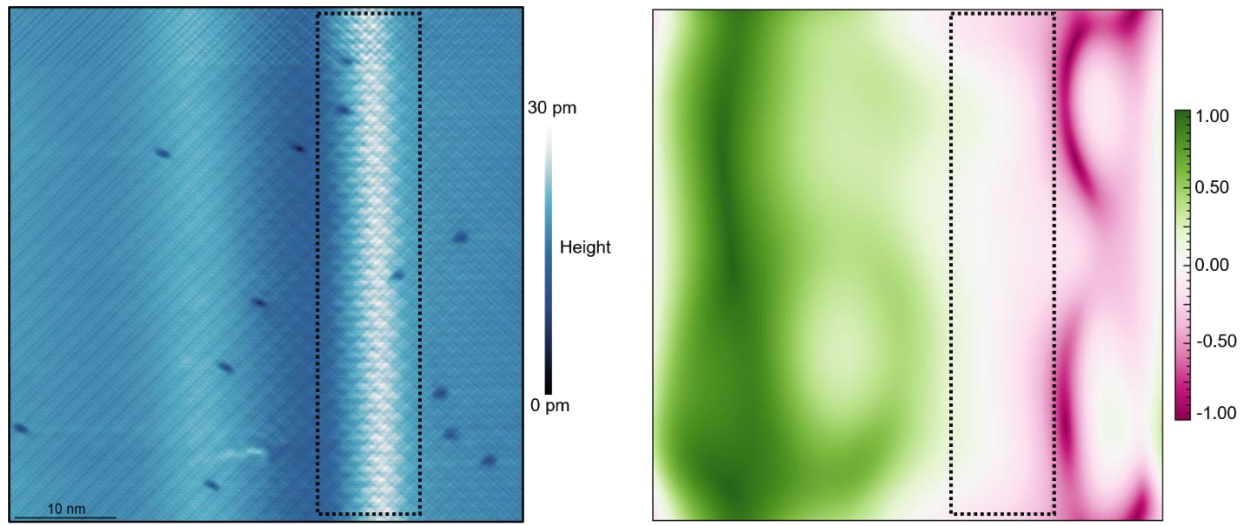

On the left, we show a 500 Å × 500 Å topography near a domain wall ($V_{Bias}$ = -60 mV, $I_{Set}$ = 60 pA). To the right, we show CDW amplitudes near the domain wall. Here, +1.00 indicates one of the CDW directions is dominant, -1.00 indicates that the perpendicular direction CDW dominates, and a value of 0.00 indicates that both CDWs are equally strong. As shown, at the domain wall, marked by the dotted black rectangles, both CDWs have equal amplitudes. All data shown in this figure was taken at 4 K.

**Supplementary Fig. 14 | Temperature-Dependent Topographies with Spin-Polarized Cr Tip**

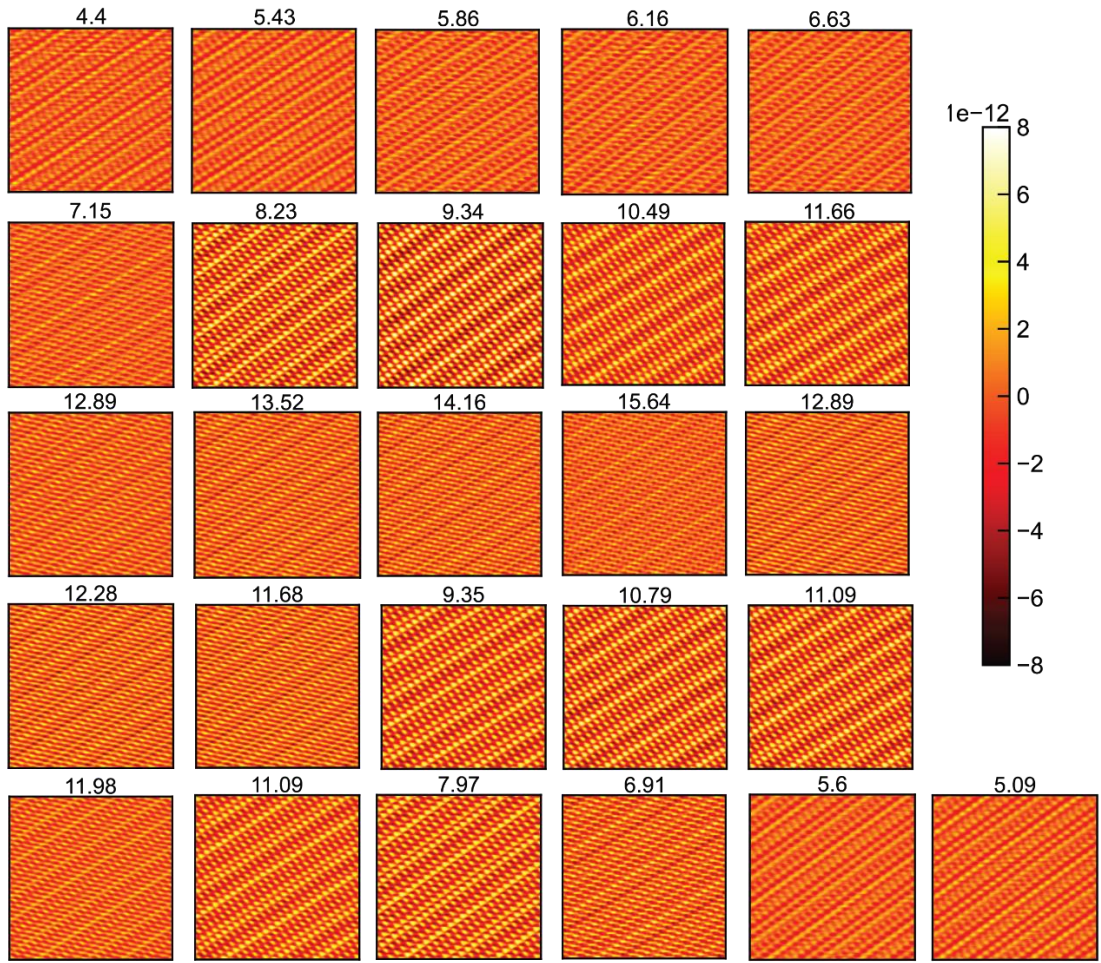

100 Å x 100 Å topographies ($V_{Bias}$ = -400 mV, $I_{Set}$ = 60 pA) from one temperature-dependent series of measurements with a common color scale, showing a clear increase in contrast at temperatures between 8 K and 12 K. Slightly more area is imaged at higher temperatures due to piezo expansion. Images are arranged in the order the data was taken to illustrate the repeatability of the contrast change. Units for temperature are K and height-scale units are m. These scans were used to generate the graph of CDW FFT peak intensity vs. temperature shown in Fig. 4a. The Fourier transforms in Supplementary Fig. 15 are from a subset of the scans taken in this run.

**Supplementary Fig. 15 | Cr Tip Temperature Dependence Measurements**

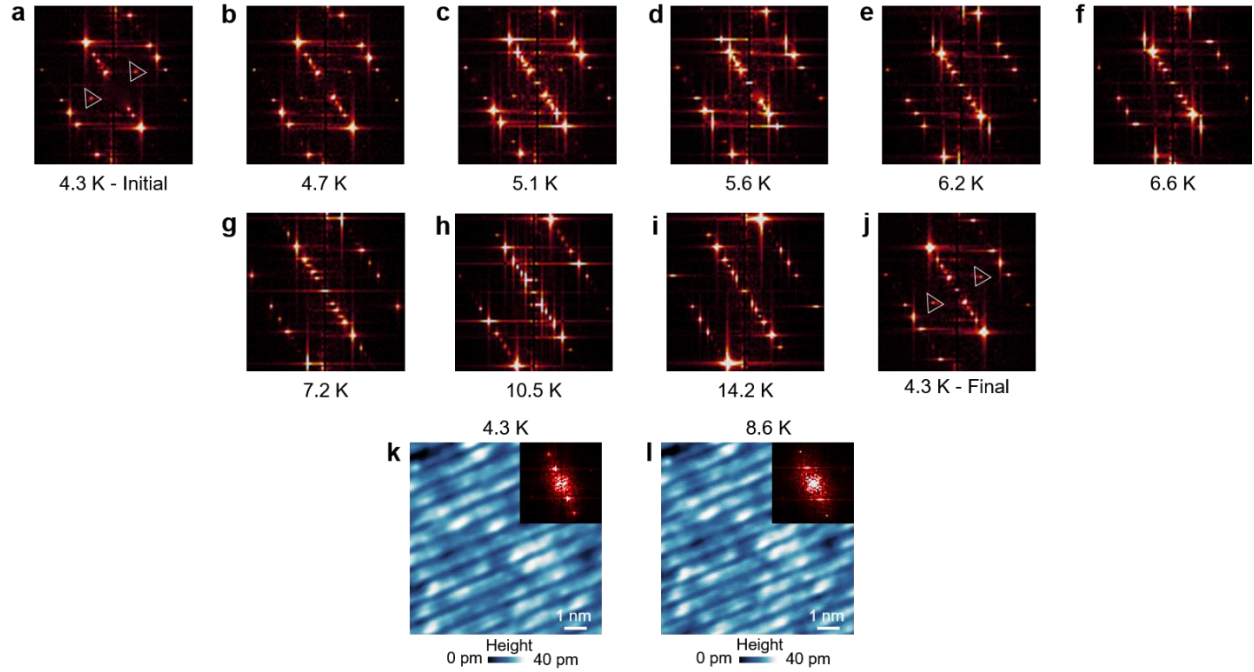

**a-j** Fourier transforms of GdTe$_3$ topographies shown in Supplementary Fig. 14 using a Cr tip. Scan were obtained at V$_{Bias}$ = -400 mV and I$_{Set}$ = 60 pA. The antiferromagnetic peaks are marked by white triangles, and show a decreasing intensity with increasing temperature, disappearing above 7.2 K. **k-l** Topographies and Fourier transforms of Fe$_{1.03}$Te with a Cr tip, at 4.3 K and 8.6 K. Scans are on the same 85 Å × 85 Å area; V$_{Bias}$ = -50 mV and I$_{Set}$ = 180 pA. The stripe antiferromagnetism persists at elevated temperatures showing that the observed disappearance of antiferromagnetism in GdTe$_3$ by ~7.5 K is not due to tip effects.

**Supplementary Fig. 16 | Cr Tip CDW Fourier Peak Temperature Dependence Repeatability**

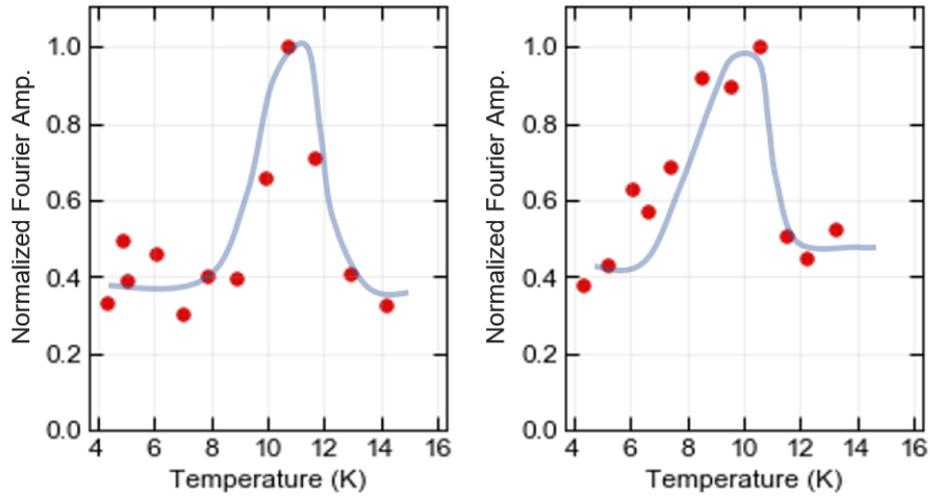

CDW Fourier peak intensities measured with a Cr-tip as a function of temperature on two different GdTe$_3$ samples with two different spin-polarized Cr-tips; y-axis normalization is to the maximum Fourier amplitude. These tips and samples are distinct from a third tip and sample used for obtaining spin-polarized data shown in Figs. 3-4. The temperature regimes shown here match those in Fig. 4, with a clear peak in all three datasets near 10 K. The topographies used to generate the left panel were taken at $V_{Bias}$ = -120 mV and $I_{Set}$ = 60 pA; the topographies used to generate the left panel were taken at $V_{Bias}$ = -180 mV and $I_{Set}$ = 60 pA.

**Supplementary Fig. 17 | Small-Scale Temperature Dependent Topographies with a W Tip**

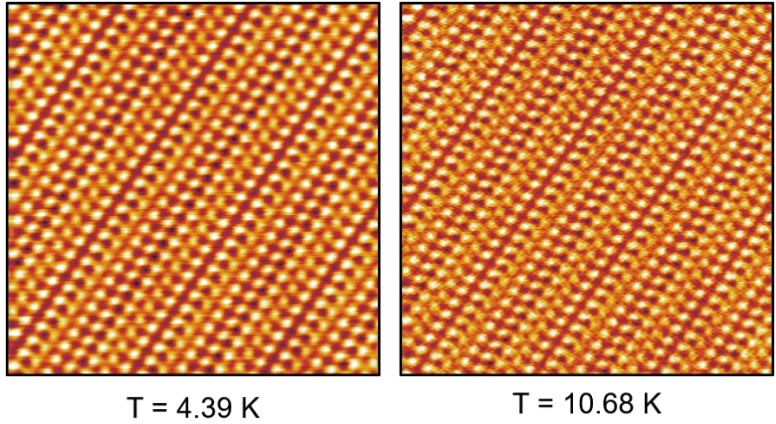

T = 4.39 K       T = 10.68 K

In contrast to the significantly different topographies taken with a Cr tip on GdTe$_3$ at ~4 K and ~10.5 K, as shown in Supplementary Fig. 14, topographies taken with a W tip show no visual changes. The scans above are 100 Å × 100 Å at $V_{Bias}$ = -120 mV and $I_{Set}$ = 240 pA.

**Supplementary Fig. 18 | Large Temperature Dependent Topographies with a W Tip**

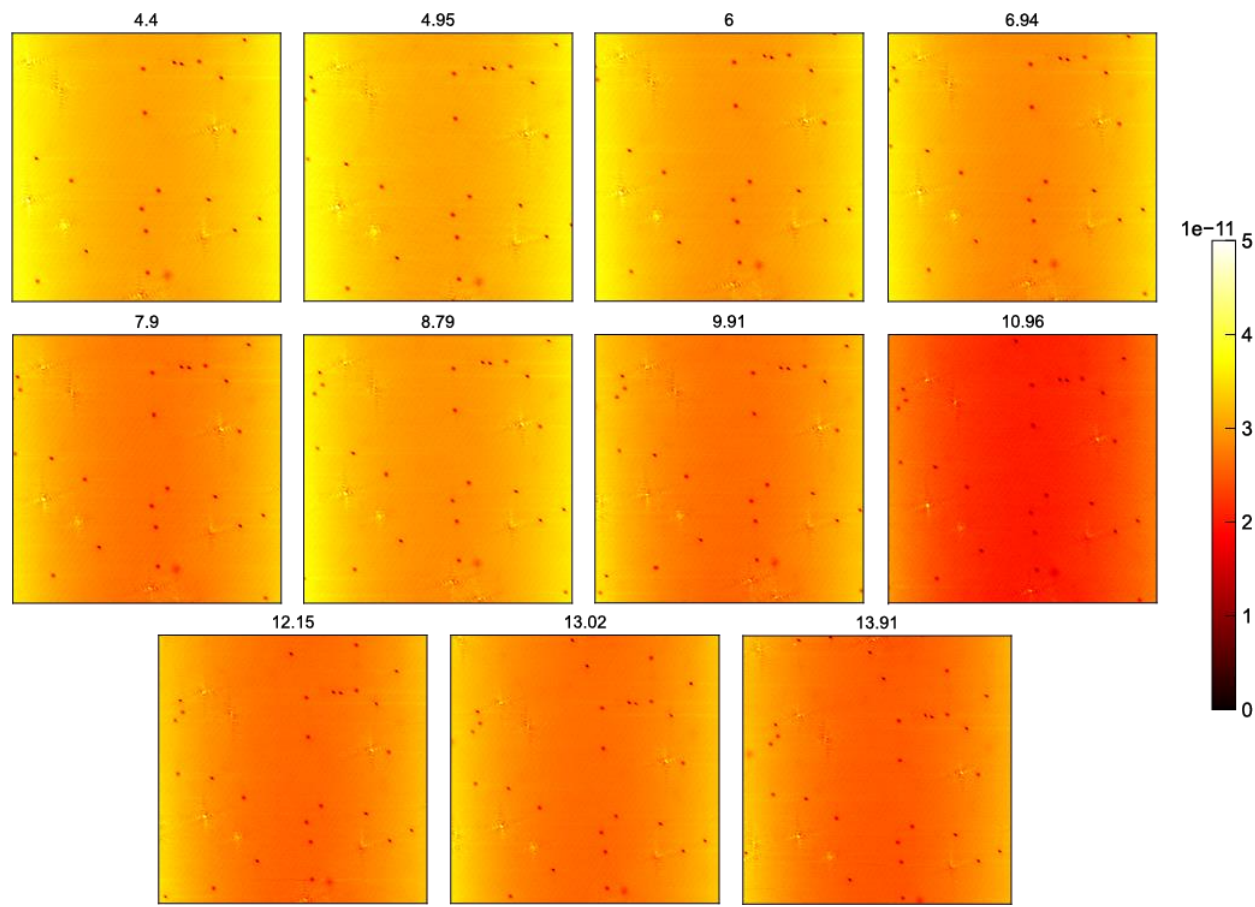

720 Å × 720 Å scans taken with a W tip at $V_{Bias}$ = -240 mV and $I_{Set}$ = 240 pA and plotted on the same color scale. These scans were used to generate the graph of CDW FFT peak intensity vs. temperature shown in Fig. 4b. The temperatures are in units of K, and the heights are in units of m.

**Supplementary Fig. 19 | W Tip Topography Bias Dependence**

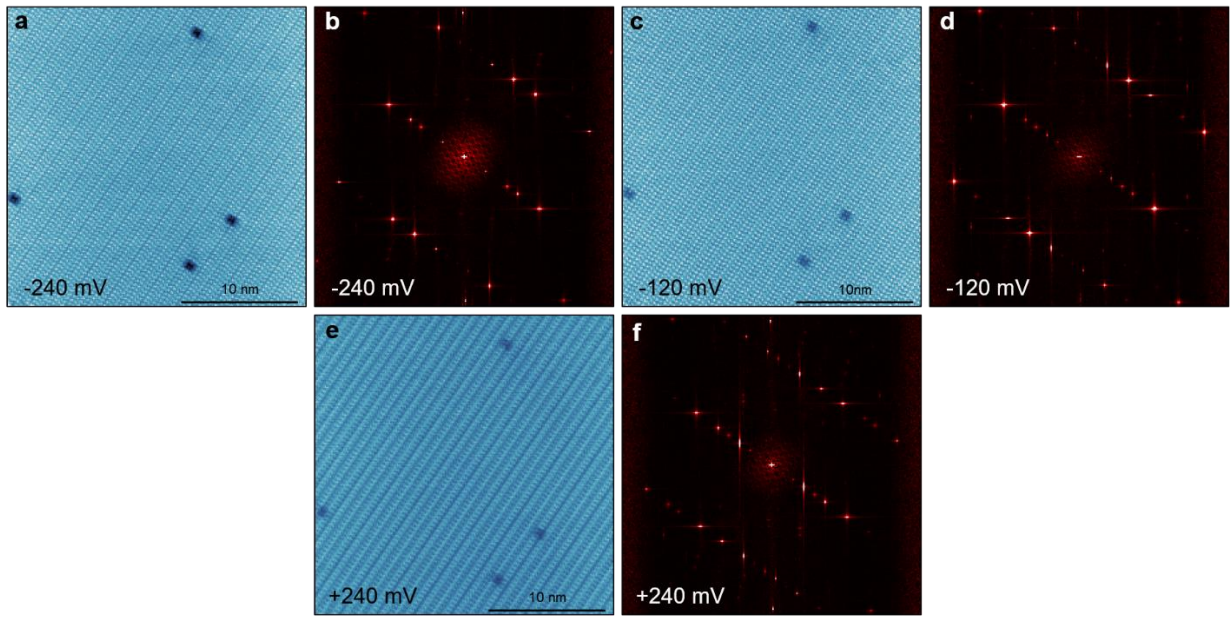

**a-f** Bias-dependent 250 Å × 250 Å scans on the same area of a GdTe$_3$ sample with a W tip, along with their Fourier transforms, at I$_{Set}$ = 240 pA. When scanning GdTe$_3$ with a W tip, the transverse peaks assigned to antiferromagnetic ordering are not seen at any bias. All data shown in this figure was taken at 4 K.

**Supplementary Fig. 20 | Robustness of Cr Tip Spin-Polarization**

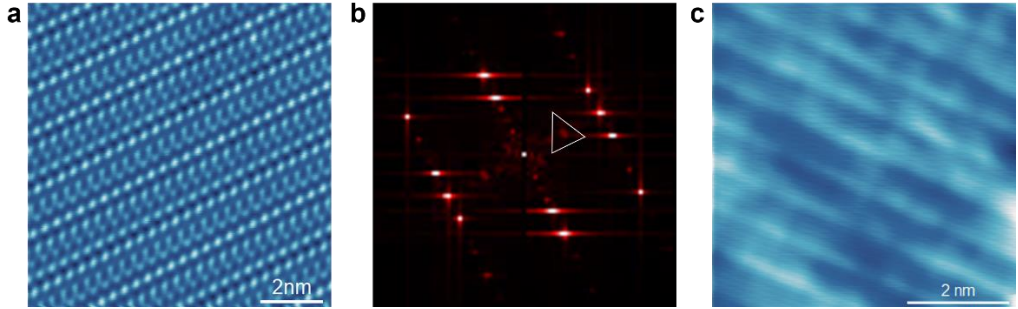

**a** Cr-tip scan of a GdTe$_3$ sample at 4 K, at $V_{Bias}$ = -60 mV and $I_{Set}$ = 480 pA; scan size is 100 Å × 100 Å. A linear plane was subtracted from the topography before the Fourier transform. **b** Raw Fourier transform of scan in **a** showing clear antiferromagnetic peaks; one of the peaks is marked by the white triangle. **c** Scan of Fe$_{1.03}$Te directly following the measurement in **a**. The tip is approached onto Fe$_{1.03}$Te at a low setpoint of 10 pA, and before any tip changes, the scan shown is taken, showing clear bicollinear antiferromagnetic stripes. The parameters for this scan are the same as those in **a**, $V_{Bias}$ = -60 mV and $I_{Set}$ = 480 pA and scan size is 60 Å × 60 Å.